\documentclass[
pre,
aps,
twocolumn,
superscriptaddress,
longbibliography,
floatfix,
notitlepage]{revtex4-2}
\usepackage[export]{adjustbox}
\usepackage[utf8x]{inputenc}
\usepackage[english]{babel}
\usepackage[T1]{fontenc}
\usepackage{lmodern}
\usepackage{epstopdf}
\usepackage{amsmath,amsfonts,amssymb,amsthm,bm,bbm,times,dcolumn}
\usepackage[normalem]{ulem}
\usepackage{microtype}
\usepackage{braket}
\usepackage{gensymb} % for example \degree symbol can be used!
\usepackage{physics}
\usepackage[title]{appendix}

\usepackage[colorlinks={true}, citecolor={blue}, filecolor={blue}, linkcolor={blue}, urlcolor={blue}]{hyperref}
\usepackage{graphicx,color}
\usepackage[caption=false]{subfig}

\begin{document}

\title{Conditions on detecting tripartite entangled state in psychophysical experiments}
\author{Lea Gassab}
\email{lgassab20@ku.edu.tr}
\author{Ali Pedram}
\email{apedram19@ku.edu.tr}
\affiliation{Department of Physics, Koç University, Sariyer 34450, Istanbul, T\"urk{\.i}ye}
\author{\"Ozg\"ur E. M\"ustecapl\ifmmode \imath \else \i \fi{}o\ifmmode \breve{g}\else \u{g}\fi{}lu}
\email{omustecap@ku.edu.tr}
\affiliation{Department of Physics, Koç University, Sariyer 34450, Istanbul, T\"urk{\.i}ye}
\affiliation{T\"UBITAK Research Institute for Fundamental Sciences, Gebze 41470,  T\"urk{\.i}ye}
\affiliation{Faculty of Engineering and Natural Sciences, Sabanci University, Tuzla 34956, Istanbul, T\"urk{\.i}ye}

\date{\today}
\pagebreak

%\keywords{Keyword1, Keyword2, Keyword3}

\begin{abstract}
This paper explores the sensitivity of the human visual system to quantum entangled light. We examine the possibility of human subjects perceiving multipartite entangled state through psychophysical experiments. Our focus begins with a bipartite entangled state to make a comparative study with the literature by taking into account additive noise for false positive on bipartite entanglement perception by humans. After that, we limit our similar investigation to a tripartite entangled state for simplicity in higher dimensions. To model the photodetection by humans, we employ the probability of seeing determined for coherently amplified photons in Fock number states, including an additive noise. Our results indicate that detecting bipartite and tripartite entanglement with the human eye is possible for a certain range of additive noise levels and visual thresholds. Finally, we discuss several alternative amplification methods.
\end{abstract}
%\begin{document}

\flushbottom
\maketitle

\section*{Introduction}

Investigating the human visual system's sensitivity to the quantum properties of light and its impact on visual functions has been a long-standing area of interest~\cite{Bouman2012,10.1085/jgp.201711970}. In-vitro studies have shown that rod cells are capable of detecting individual photons~\cite{RevModPhys.70.1027,YAU1977,https://doi.org/10.1113/jphysiol.1979.sp012715,https://doi.org/10.1113/jphysiol.1979.sp012716,https://doi.org/10.1113/jphysiol.1984.sp015518,Reingruber19378,PhysRevLett.112.213601,PhysRevLett.109.113601}. However, early psychophysical experiments in dark adaptation conditions suggested that while humans can detect a single photon in their retina, visual perception only becomes possible when a threshold of several photons is reached~\cite{10.1085/jgp.25.6.819,vanderVelden1946,Barlow:56,doi:10.1113/jphysiol.1972.sp009838,Teich:82,Prucnal1982,Teich1982}. The earliest experimental estimate of this threshold is 5-7 photons~\cite{10.1085/jgp.25.6.819}. Later studies revealed that the visual threshold and noise in the visual pathway are connected and it is crucial to take into account the dark counts, false positives and experimental uncertainties in general, in order to interpret and understand the data collected in the psychophysical experiments~\cite{Barlow:56,doi:10.1113/jphysiol.1972.sp009838,Teich:82,Prucnal1982,Teich1982,barlow1977retinal,DONNER1992853,LILLYWHITE1981291,doi:10.1146/annurev.physiol.67.031103.151256}. Recent research suggests that humans may have the ability to directly perceive a single photon with a probability greater than 50\%, and that this efficiency increases with earlier photon absorption~\cite{Tinsley2016}. These intriguing findings present the potential for the use of quantum technologies, particularly quantum metrology, in biometry~\cite{PhysRevApplied.8.044012}, the study of retinal circuits~\cite{PhysRevResearch.4.033060}, the diagnosis of visual and brain disorders~\cite{10.1093/brain/awp068,bellotti2012advanced,davodabadi2023mathematical}, for exploring the boundaries of classical and quantum theories~\cite{PhysRevA.78.052110}, and creating high-resolution images of the retina~\cite{Berchera_2019}.

Brunner et al. proposed the detection of bipartite entangled states in a Bell experiment through the use of human eyes modeled as threshold photodetectors~\cite{PhysRevA.78.052110}. However, they found that for a more accurate eye model that accounted for the smooth probability of seeing, postselection was necessary to verify the violation of the Bell inequality. Sekatski et al. demonstrated that the violation of the Bell inequality could be confirmed without postselection by amplifying one photon of an entangled pair through stimulated emission~\cite{PhysRevLett.103.113601}. Vivoli et al. created an entanglement witness to assess the ability of the human eye to detect bipartite photonic entanglement~\cite{Vivoli:16}. They proposed an interferometric setup that amplifies low photon numbers with coherent state generators before detection by the human eye. Building on Vivoli et al.'s methodology, Dodel et al. showed that it is possible to detect non-classical (sub-Poissonian) light states using human eyes as photodetectors~\cite{Dodel2017proposalwitnessing}. Sarenac et al. experimentally demonstrated that human subjects can distinguish between two polarization-coupled orbital angular momentum (OAM) states when viewing a structured light beam~\cite{doi:10.1073/pnas.1920226117}. More recently, it was demonstrated that human subjects can perceive different OAM modes created through Pancharatnam-Berry phases~\cite{Sarenac2022}.

This study aims to determine if human subjects can indirectly infer the presence of higher-dimensional entanglement, quantum entangled light through psychophysical experiments, starting with the simpler case of bipartite entangled states before considering tripartite entangled states to understand the impact of psychophysical conditions. Our work is different from the pioneering works~\cite{Dodel2017proposalwitnessing,Vivoli:16,doi:10.1073/pnas.1920226117,Sarenac2022} in several crucial aspects. First, we consider the case of higher dimensional entanglement, in particular tripartite entanglement. This requires a different type of entanglement witness as there are distinct entanglement classes, W and GHZ states, for higher order entangled states. We specifically consider W states and use a witness which is experimentally accessible for our purposes. Second, both for bipartite and higher dimensional entanglement, we take into account additional imperfections arising in psychophysical experiments. The fundamental significance of our results is that we explore and present results on the limits of using human eyes for entanglement detection. Recent studies demonstrated that humans can detect single photons and entangled photon pairs, but the question of what is the limit of humans to detect higher dimensional entanglement remained open. This question has a further practical impact as quantum advantages become most useful with scaled entanglement, which can bring unprecedented precision and perspectives in probing human visual system via psychophysical experiments.

In our work we present a detection method for entangled optical states, both bipartite and tripartite. We use the interferometry setup~\cite{Vivoli:16} and the entanglement witness~\cite{PhysRevA.101.062319} to investigate whether humans can detect bipartite entanglement. Additionally, we extend the interferometer to measure tripartite W state entanglement witness~\cite{PhysRevLett.92.087902} and explore whether human subjects can detect such multipartite entanglement.
In our photodetection scheme, we model human subjects as biological photodetectors and incorporate the noise due to photon loss in the eye and the additive noise due to false positive into the ``probability of seeing" distributions. Our results suggest that the detection of  entangled quantum light is possible for a range of noise levels and, more critically, the visual threshold. Finally, we also discuss the potential to use better amplification methods for noise reduction.

The structure of this paper is as follows:~\ref{sec:exset} describes the simulation method employed for the setups for bipartite and tripartite entangled state generation and detection. \ref{sec:Prob} outlines the approach used for the probability of seeing and the implementation of noise. \ref{sec:results} presents the results of the study. \ref{sec:amp} discusses other amplification possibilities as alternatives to coherent addition for the amplification process. \ref{sec:conclusion} provides a summary and discusses the findings presented in this paper.

%%%%%%%%%%%%%%%%%%%%%%%%%%%%%%%%%%%%%%%%%%%%

\section{Model System}
\label{sec:exset}
We focus our analysis to a Bell state for the bipartite case and a W state for the tripartite case assuming that they have been produced by the interferometric setups in \cite{Vivoli:16} and \cite{PhysRevLett.92.087902} respectively and horizontal polarization components are probed by human eyes used in place of photodetectors.

\begin{figure}[t!]
        \centering
        \captionsetup{justification=justified}
        \subfloat[]{{\includegraphics[width=\linewidth]{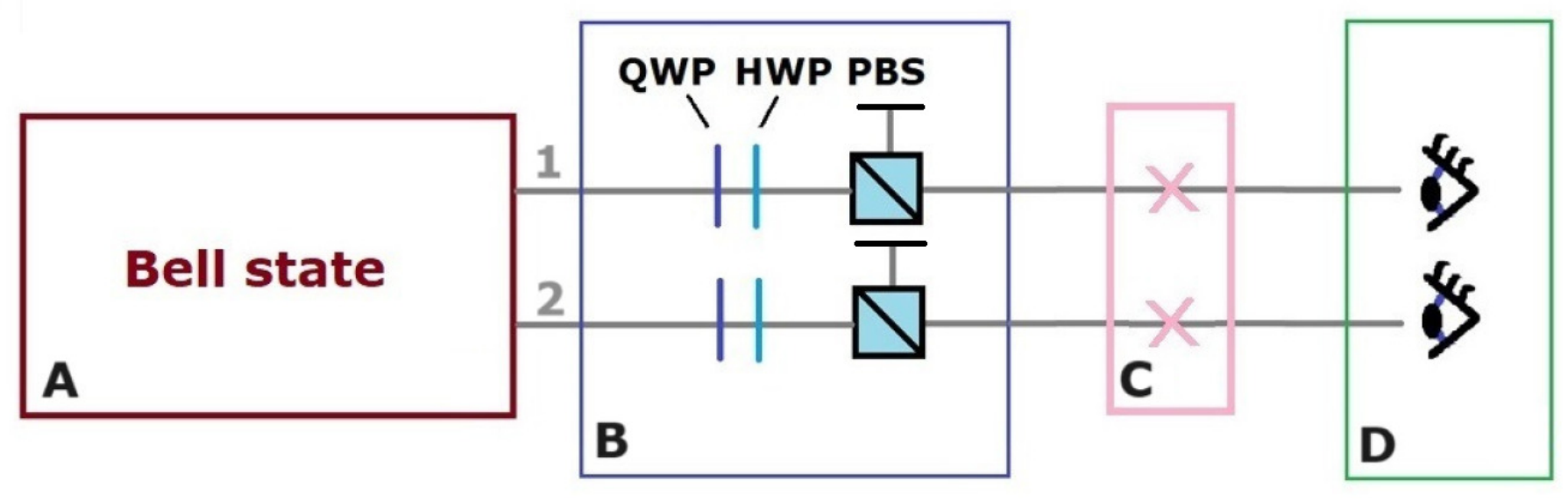}}}
        \hfil
        \subfloat[]{{\includegraphics[width=\linewidth]{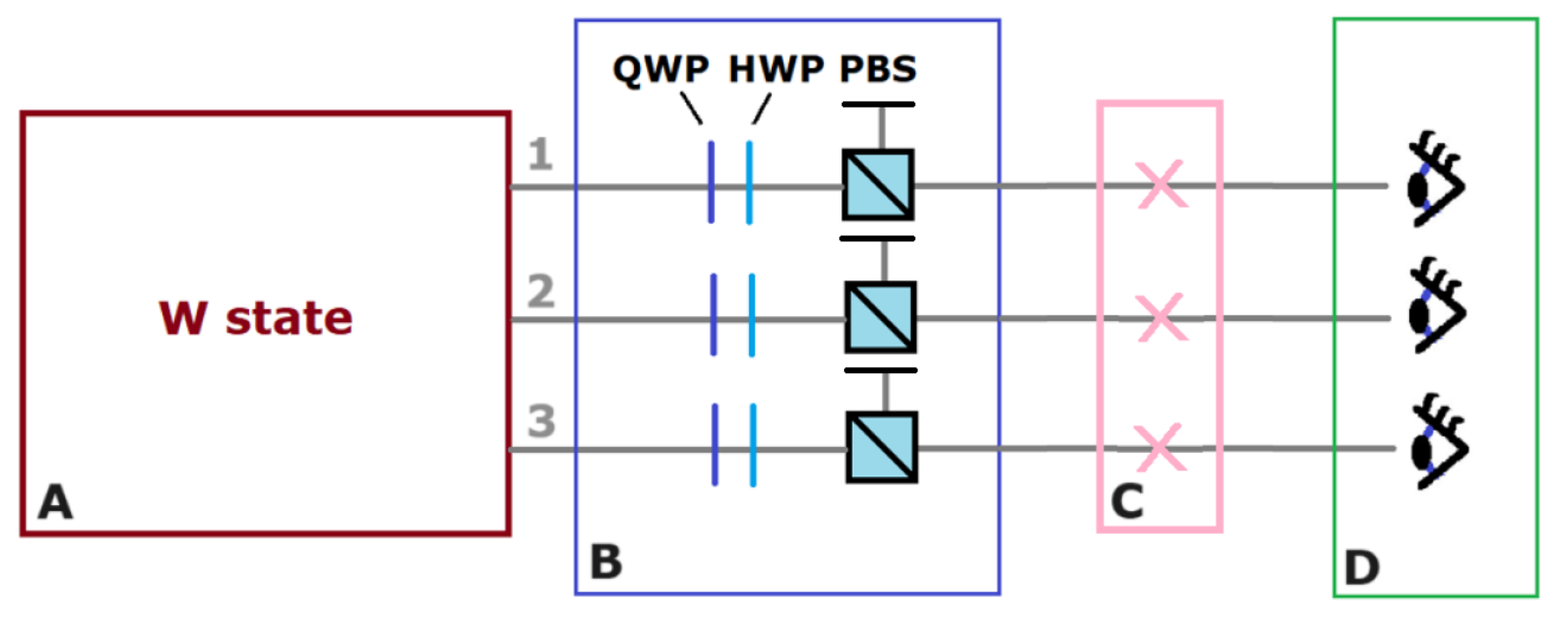}}}
        \caption{The red boxes (labeled by $A$) are schematic representation of the interferometric entanglement generation setups ~\cite{Vivoli:16,PhysRevLett.92.087902}. Upper (a) and lower (b) panels are respectively for the bipartite and tripartite cases. The blue boxes (labeled by $B$) represent the polarization analyzers complexes, which includes Quarter Wave Plate (QWP), a Half Wave Plate (HWP) and a Polarized Beam Splitter (PBS). The pink boxes (labeled by $C$) represent the coherent amplification processes. The green boxes (labeled by $D$) represent the detection part of the horizontal polarization by the human eye. A horizontal black line appears at one of the outputs of the PBS to illustrate the beam being blocked. The numbers 1,2 and 3 correspond to the branches in the interferometric setup.}
       \label{Fig1}
\end{figure}

The studies of bipartite photonic entanglement and tripartite photonic entanglement start respectively with the states,
\begin{align}
\ket{\psi_1}&=\frac{1}{\sqrt{2}}(\ket{HV} + \ket{VH});\label{eq1}\\
\ket{\psi_2}&=\frac{1}{\sqrt{3}}(\ket{HHV} + \ket{HVH} + \ket{VHH}),\label{eq2}
\end{align}
where $H$ stands for horizontal polarization and $V$ for vertical polarization \cite{Vivoli:16,PhysRevLett.92.087902}.
We choose to represent the starting states in terms of creation operator, $\hat{a}^{\dag}$, applied to the vacuum as follows:
\begin{align}
\ket{\psi_1} &= \frac{1}{\sqrt{2}}(\hat{a}^{\dag1}_H \hat{a}^{\dag2}_V + \hat{a}^{\dag1}_V \hat{a}^{\dag2}_H) \ket{0},\label{eq3}\\
\ket{\psi_2} &= \frac{1}{\sqrt{3}}(\hat{a}^{\dag1}_H \hat{a}^{\dag2}_H \hat{a}^{\dag3}_V + \hat{a}^{\dag1}_H \hat{a}^{\dag2}_V \hat{a}^{\dag3}_H + \hat{a}^{\dag1}_V \hat{a}^{\dag2}_H \hat{a}^{\dag3}_H) \ket{0},\label{eq4}
\end{align}
with the superscripts (1,2,3) corresponding to the branches in the interferometric setup (see~\ref{Fig1}) and the subscripts ($H$,$V$) corresponding to the horizontal and vertical polarization.
As the human visual threshold cannot resolve single photons, or two or three-photons, after the photons go through the polarization analyzers complex, a coherent light is mixed with the entangled photons so that the mean photon number is coherently amplified above the visual threshold.
Theoretically, this amplification means applying to the horizontal polarized photons a displacement operator,
\begin{equation}\label{eq5}
\hat{D}(\alpha) = \mathrm{e}^{\alpha \hat{a}^{\dag} - \alpha^{*} \hat{a}},
\end{equation}
where $\alpha$ is the displacement parameter; $\hat{a}^{\dag}$ and $\hat{a}$ are respectively the photon creation and annihilation operators.

The measurement in the vacuum state and single-photon subspace yields two outcomes (termed ``no see" and ``see") and is represented as a positive operator-valued measure (POVM) on the Bloch sphere. The measurement direction depends on the displacement's amplitude and phase, enabling effective measurements, particularly along the x-axis. The method capitalizes on the partial overlap in photon number space due to the displacement operation.
In our case the main role of the displacement operator is to increase the number of photons to a level above the human visual threshold when a photon is present. Conversely, in the absence of photons, the amplification remains below the threshold. This approach effectively boosts the contrast between ``yes" and ``no" cases, making photon detection more robust and accurate.

\begin{figure*}[t!]
        \centering
        \captionsetup{justification=justified}
        \includegraphics[width=\linewidth]{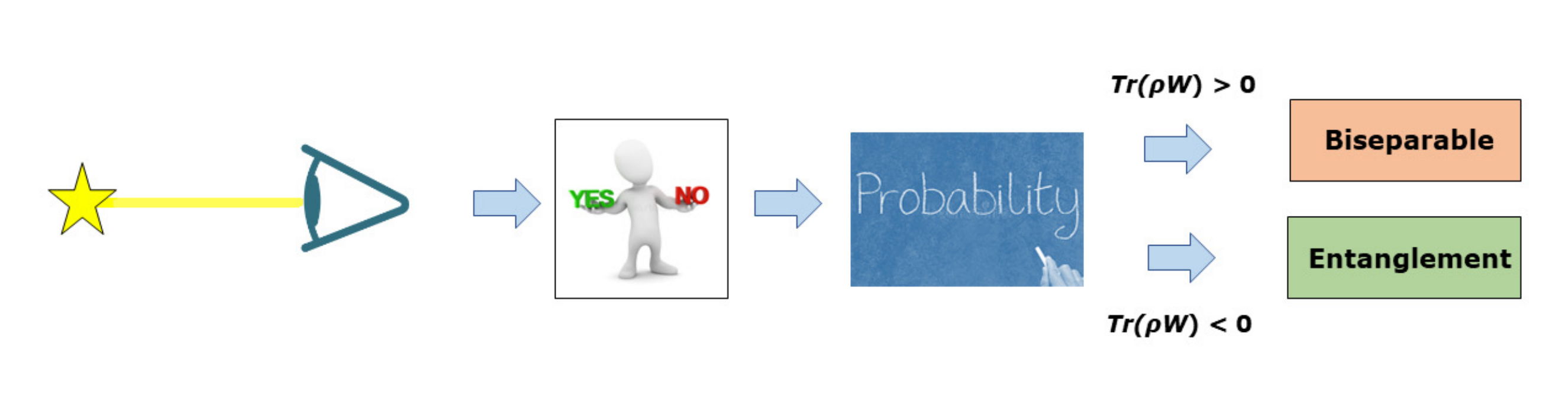}
        \centering
        \caption{Simplified representation of the method used in the study. The light source is represented by a yellow star. The light coming into the human eye is represented by the solid yellow line. Then, the human subject answers by yes or no according to whether he/she sees or does not see the light. After $N$ simulations of the experiment, we calculate the probabilities needed in the different witnesses given in Eq.~(\ref{eq10}) and Eq.~(\ref{eq11}). From these probabilities, we can calculate the expectation value of the entanglement witness and conclude if the human subject can differentiate between entangled and non-entangled light or multipartite and biseparable light.}
       \label{Fig2}
\end{figure*}
The polarization analyzers complex consists of a quarter wave plate (QWP), a half wave plate (HWP) and a polarized beam splitter (PBS)~\cite{Hou:15}. Considering the unitary transformation operator as defined in Eq.~(2) of Hou et al.~\cite{Hou:15},
\begin{equation}\label{eq6}
U(\theta,\delta)=
\begin{pmatrix}
\cos^2{\theta} + \mathrm{e}^{i\delta} \sin^2{\theta} & \frac{1}{2}(1-\mathrm{e}^{i\delta} \sin{2\theta}) \\
\frac{1}{2}(1-\mathrm{e}^{i\delta} \sin{2\theta}) & \sin^2{\theta} + \mathrm{e}^{i\delta} \cos^2{\theta}
\end{pmatrix},
\end{equation}
the QWP is represented by $U(q,\frac{\pi}{2})$ and the HWP by $U(p,\pi)$ where $q$ and $p$ are real numbers.
In each interferometric branch the creation operators for horizontal and vertical polarization are modified by the application of the unitary operators,
\begin{equation}\label{eq7}
U(q,\frac{\pi}{2})~U(p,\pi)
\begin{pmatrix}
\hat{a}^{\dag}_H \\
\hat{a}^{\dag}_V
\end{pmatrix}=
\begin{pmatrix}
\hat{a}_H^{\dag'} \\
\hat{a}_V^{\dag'}
\end{pmatrix},
\end{equation}
with $\hat{a}_H^{\dag'}$ and $\hat{a}_V^{\dag'}$ being the modified creation operators after going through the polarization analyzer.
Equivalently, as explained by Hou et al.~\cite{Hou:15}, the polarization analyzers complex transforms the  polarization operator $\sigma_z$ as follows:
\begin{equation}\label{eq8}
\sigma_z \mapsto r(q,p).\sigma,
\end{equation}
in which $\sigma=(\sigma_x,\sigma_y,\sigma_z)$ is the row vector of Pauli spin-$1/2$ operators,  and $r(q,p)$ is a column vector given by
\begin{equation}\label{eq9}
r(q,p)=
\begin{pmatrix}
\sin(2q)\cos(4p-2q) \\
\sin(4p-2q) \\
\cos(2q)\cos(4p-2q)
\end{pmatrix},
\end{equation}
where $p$ and $q$ are chosen in such a way to be able to obtain the appropriate polarization analyzer setting. For example, if we need to measure $\sigma_y$, then we take $q=0$ and $p=\pi/8$ which gives
\begin{equation}
r(0,\pi/8)=
\begin{pmatrix}
0 \\
1 \\
0
\end{pmatrix}.
\end{equation}

In a real experimental scenario, the PBS in the polarizer analyzer complex separates the vertically and horizontally polarized photons as illustrated by the box B in~\ref{Fig1}. At this point, it is worth noting that, when performing computational simulations, the polarized photons are already separated by the two tensor products of the creation operator representing the horizontal and vertical polarization in each interferometric branch.

Next, the horizontal polarization are amplified (Box C of~\ref{Fig1}), and transmitted to and detected by the human eye (Box D of~\ref{Fig1}). The schematic representation of the full setup is shown in~\ref{Fig1}.

To decide whether or not our state is entangled (multipartite entangled), we need an entanglement witness. In the quantum information and quantum optics literature there exists a plethora of entanglement witnesses. We choose two specific witnesses that can be implemented in an experimental setup, one for the bipartite entanglement witness~\cite{PhysRevA.101.062319}, and the other for the tripartite entanglement witness~\cite{PhysRevLett.92.087902}. If the expectation value of either of the entanglement witness is strictly negative, we can deduce that we have an entangled or multipartite entangled state.

The entanglement witness for bipartite entanglement and multipartite entanglement can be decomposed into a number of local von Neumann measurements~\cite{PhysRevA.101.062319,PhysRevLett.92.087902},

\begin{align}
    W_1&= \frac{1}{4}[ \mathbbm{1}  \otimes  \mathbbm{1}  + \sigma_z \otimes \sigma_z - \sigma_y \otimes \sigma_y -\sigma_x \otimes \sigma_x]; \label{eq10}\\
    \begin{split}
        W_2&= \frac{1}{16}[ 6\cdot\mathbbm{1}  \otimes  \mathbbm{1} \otimes  \mathbbm{1}  \\
        &+4\cdot \sigma_z     \otimes \sigma_z \otimes \sigma_z - 2\cdot (\sigma_y \otimes \sigma_y \otimes \mathbbm{1} \\
        &+ \sigma_y \otimes  \mathbbm{1}     \otimes \sigma_y + \mathbbm{1} \otimes  \sigma_y \otimes \sigma_y)  \\
        &  -(\sigma_x + \sigma_z) \otimes (\sigma_x + \sigma_z) \otimes (\sigma_x + \sigma_z)  \\
        &  - (\sigma_z - \sigma_x) \otimes (\sigma_z - \sigma_x) \otimes (\sigma_z - \sigma_x) ].\label{eq11}
    \end{split}
\end{align}
These measurements correspond to different combinations of Pauli matrices. Therefore, the witnesses can be simply expressed as,
\begin{equation}
    \label{eq12}
    W=\sum_{k=1}^{S} M_k,
\end{equation}
where the observable $M_k$ corresponds to the $k^{\mathrm{th}}$ combination of Pauli matrices appearing in the witnesses and $S$ denotes the number of different settings needed. Actually, the different combination of Pauli matrices are generated thanks to the polarization analyzer as presented in Eq.~(\ref{eq8}). In our analysis, we need $S=3$ different settings for Bell state detection and $S=5$ different settings for W state. For example, when looking at the witness for Bell state in Eq.~(\ref{eq10}), we need three setups of the polarization analysers to recover the witness, which means having in Eq.~(\ref{eq8}) $q=0$, $p=0$ for $\sigma_z \otimes \sigma_z$,  $q=0$, $p=\pi/8$ for $\sigma_y \otimes \sigma_y$ and $q=\pi/8$, $p=0$ for $\sigma_x \otimes \sigma_x$. For each setting, we calculate the probabilities of the different states before the detection and we generate these states in the simulation according to their probabilities.

Then, the detection is realized by a human eye or by a perfect photodetector.
\begin{itemize}
  \item If the human receives an $H$ polarized photon, the human will see light with the probability, $P_{\mathrm{see}}$ given in Eq.~(\ref{eq20}).
  \item If the human eye does not receive any photon, (that is, the photons were in $V$ polarization and hence not transmitted to the eye) then the probability of seeing will not be zero due to the amplification process and the person will see light with a different probability $P_{\mathrm{see}}$.
  \item If we use a perfect photodetector in place of a human eye, the probability of detection will be 1 for $H$ polarization and 0 for $V$ polarization.
\end{itemize}
Thus, according to the simultaneous responses of the human eyes or the output of the photodetectors, we can construct four different probabilities for the Bell state,
\begin{equation*}
P(HH), P(HV), P(VH), P(VV)
\end{equation*}
corresponding to the photonic states with the polarizations,
\begin{equation*}
HH, HV, VH, VV,
\end{equation*}
and eight different probabilities for the W state,
\begin{align*}
&P(HHH), P(HHV), P(HVH), P(HVV),\\
& P(VHH), P(VHV), P(VVH), P(VVV)
\end{align*}
corresponding to the photonic states with the polarizations,
\begin{equation*}
HHH, HHV, HVH, HVV, VHH, VHV, VVH, VVV.
\end{equation*}
From these probabilities, for different setup, we can calculate the expectation value of the witness after $N$ simulations,
\begin{equation}\label{eq13}
\begin{split}
\left\langle W \right\rangle &= \sum_k Tr(M_k \rho)\\
&= \sum_k \sum_{l_1,...,l_n = V,H} d_{l_1,...,l_n}^{(k)} P^{(k)}(l_1,...,l_n),
\end{split}
\end{equation}
where $\rho$ is the density matrix of the state~\cite{PhysRevLett.92.087902}. $l_n$ corresponds to the polarization of the photon in the interferometric branch $n$ such that $n=2$ for the Bell state and $n=3$ for the W state. $d_{l_1,...,l_n}^{(k)}$ are real weight coefficients taken from the witnesses (\ref{eq10}) and (\ref{eq11}) and $k$ is the setting of the polarization analyzer. $P^{(k)}(l_1,...,l_n)$ are the probabilities of the different possible outputs given by the human or photodetector. In our setup, we employ perfect photodetectors. However, it is important to acknowledge the quantum efficiency of typical photodetectors in practical scenarios, where their performance remains consistently high \cite{shen2022near}. Although there might be minor detrimental effects on the overall witness measurement, these are anticipated to be minimal.

Finally, while the human eye detects photons like a threshold photodetector, it is through analysis of the responses to a large number of photon counting events that we can deduce the existence of either bipartite or multipartite entanglement. Indeed after calculating the probabilities using the counting statistics, given our knowledge of the source of light, we can calculate the expectation value of the entanglement witness. If this value is negative ($\left\langle W \right\rangle < 0$), we can conclude that the photonic source contains bipartite or tripartite entanglement. The general strategy is summarized in~\ref{Fig2}.

%%%%%%%%%%%%%%%%%%%%%%%%%%%%%%%%%%%%%%%%%%%%%%

\section{Probability of seeing and noise}
\label{sec:Prob}

\begin{figure}[t!]
        \centering
        \captionsetup{justification=justified}
        \includegraphics[width=\linewidth]{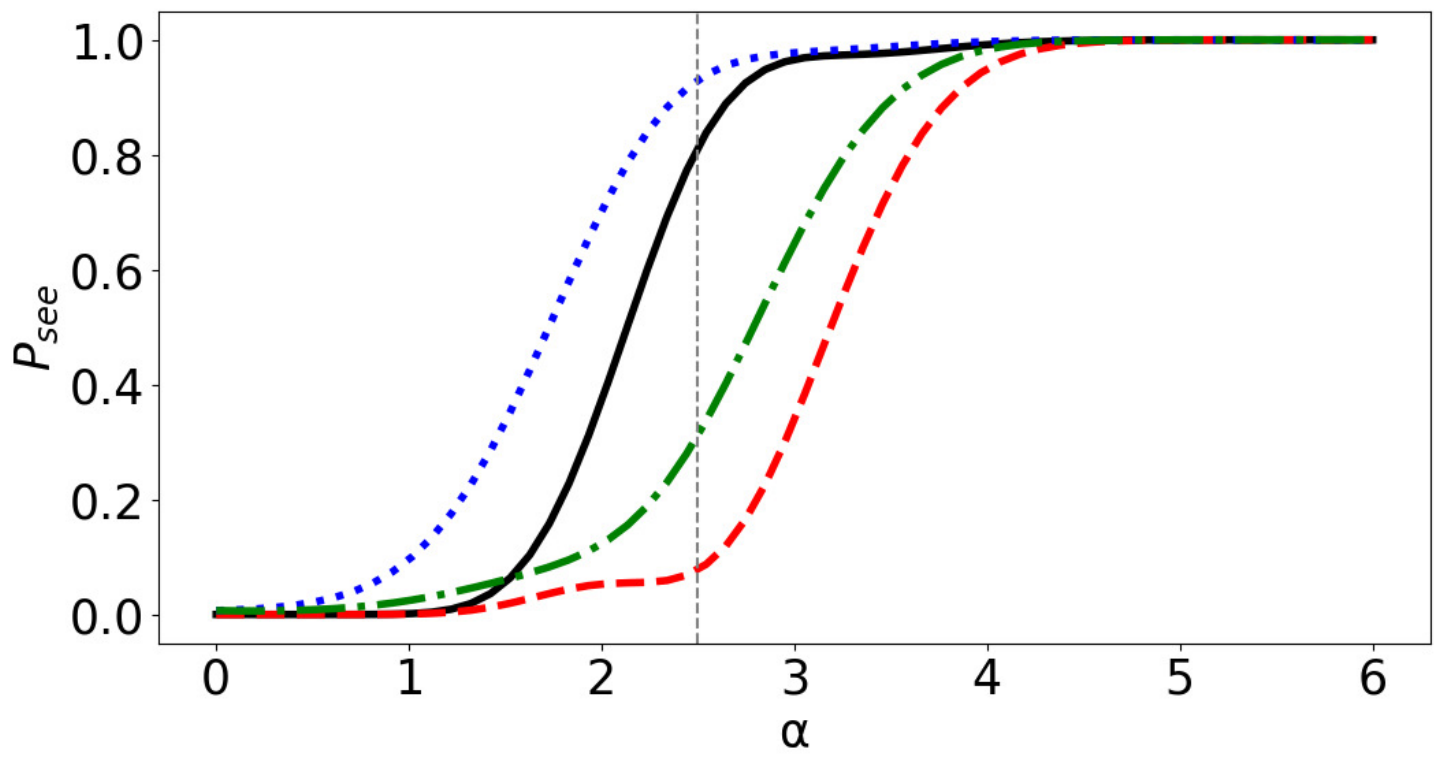}
        \caption{Probability of seeing $P_{\mathrm{see}}$ (Eq.~(\ref{eq20})) as a function of the coherence parameter $\alpha$ of the coherent amplification. The solid black line represents $P_{\mathrm{see}}$ for the state  $\ket{1'}$ and the dashed red line represents $P_{\mathrm{see}}$ for the state $\ket{0'}$. The dotted blue line represents $P_{\mathrm{see}}$ for the state  $\ket{1'}$ with additive noise $D=2$ and the dashed dotted green line represents $P_{\mathrm{see}}$ for the state $\ket{0'}$ with additive noise $D=2$. The threshold $K$ is taken as 7 and $\eta=0.08$ as in Vivoli et al.~\cite{Vivoli:16}. The dashed vertical gray line is given for $\alpha=2.5$.}
       \label{Fig3}
\end{figure}
To simulate the human response, we need to define the probability of seeing. The probability of seeing will depend on the input light coming to the eye.
The occurrence of a no-seeing event with unit efficiency can be associated with
\begin{equation}
\hat{P}_{\mathrm{ns}}^{K,\eta=1}=\sum_{m=0}^{K-1} \ket{m}\bra{m},
\end{equation}
where $K$ is the human threshold and $\eta$ is the quantum efficiency. To model non-unit efficiency, we can introduce a beam-splitter interaction between the detected mode $A$ and an unoccupied ancillary mode $C$. These modes are described by the bosonic operators $a$, $a^{\dagger}$, and $c$, $c^{\dagger}$ respectively. The beam splitter operator ($U_{BS})$ can be expressed as follows,

\begin{equation}
U_{BS}\ket{0}_c=C_L=\mathrm{e}^{\tan\gamma a c^{\dagger}}\mathrm{e}^{\ln(\cos\gamma)a^{\dagger}a}\ket{0}_c,
\end{equation}
with $\cos^2\gamma=\eta$.
The no-seeing operator is defined as,

\begin{equation}
\hat{P}_{\mathrm{ns}}^{K,\eta=1}=C_L^{\dagger}\sum_{m=0}^{K-1} \ket{m}\bra{m}C_L.
\end{equation}
The no-seeing probability is eventually given by the operator,
\begin{equation}\label{eq14}
\hat{P}_{\mathrm{ns}}^{K,\eta}=\frac{\eta^{K}}{(K-1)!}\frac{d^{K-1}}{d(1-\eta)^{K-1}} \left[ \frac{(1-\eta)^{\hat{a}^{\dag}\hat{a}}}{\eta} \right],
\end{equation}
where $\frac{d^{K-1}}{d(1-\eta)^{K-1}}$ is the $(K-1)^{th}$ derivative with respect to $(1-\eta)$.
Then, the probability of no-seeing after coherent amplification is calculated as follows,
\begin{align}
P_{\mathrm{ns}}^{K,\eta}&=\bra{n}\hat{D}(\alpha)^{\dag} \hat{P}_{ns}^{K,\eta} \hat{D}(\alpha)\ket{n}\label{eq15}\\
&= \frac{\eta^{K}}{(K-1)!}\frac{d^{K-1}}{d(1-\eta)^{K-1}} \left[ \frac{\mathrm{e}^{-\eta|\alpha|^2}p^{\alpha}_n(\eta) }{\eta}\right],\label{eq16}
\end{align}
where
\begin{equation}\label{eq17}
p^{\alpha}_n(\eta)=\bra{n}\mathrm{e}^{-\eta \alpha \hat{a}^{\dag}}(1-\eta)^{\hat{a}^{\dag}\hat{a}}\mathrm{e}^{-\eta\alpha^{*}\hat{a}}\ket{n},
\end{equation}
with $n$ being the number of horizontal polarized photon in the interferometric branch before the amplification process (0 or 1 in this study). Although eye-based measurements are incapable of assessing the z-axis, they exhibit a level of comparability to single photon detectors when it comes to conducting measurements along the x-axis. When the state is amplified via coherent amplification, human or any detector is no longer interrogated with respect to amplified $\ket{0}$ and $\ket{1}$ states but with respect to amplified $\ket{0+1}$ and $\ket{0-1}$  states. So, instead of differentiating between the states $\ket{1}$ and $\ket{0}$ we differentiate between the states,
\begin{align}
\ket{1'}&=\frac{1}{\sqrt{2}}(\ket{0}+\ket{1});\\
\ket{0'}&=\frac{1}{\sqrt{2}}(\ket{0}-\ket{1}).
\end{align}
This improves the distinguishability between these two states by the human eye. Using the states $\ket{0'}$ and $\ket{1'}$ for $\ket{n}$ in Eq.~(\ref{eq17}), we can compute the probability of Eq.~(\ref{eq15}) and use
\begin{equation}\label{eq20}
P_{\mathrm{see}}=1-P_{\mathrm{ns}}^{K,\eta}
\end{equation}
as the probability of seeing in our simulation. In Figure~\ref{Fig3}, the probability of seeing for state $\ket{1'}$ corresponding to presence of a horizontal polarized photon and $\ket{0'}$ corresponding to absence of photon are represented. The threshold $K$ is varied up to $7$ and the efficiency $\eta$ is equal to $0.08$~\cite{Vivoli:16}. Based on Figure~\ref{Fig3}, we choose for the rest of the study the displacement parameter $\alpha=2.5$ in order to get the best distinguishability between the states $\ket{0'}$ and $\ket{1'}$.

\begin{figure*}[t!]
		\centering
  \captionsetup{justification=justified}
		\subfloat[]{\includegraphics[width=0.45\linewidth]{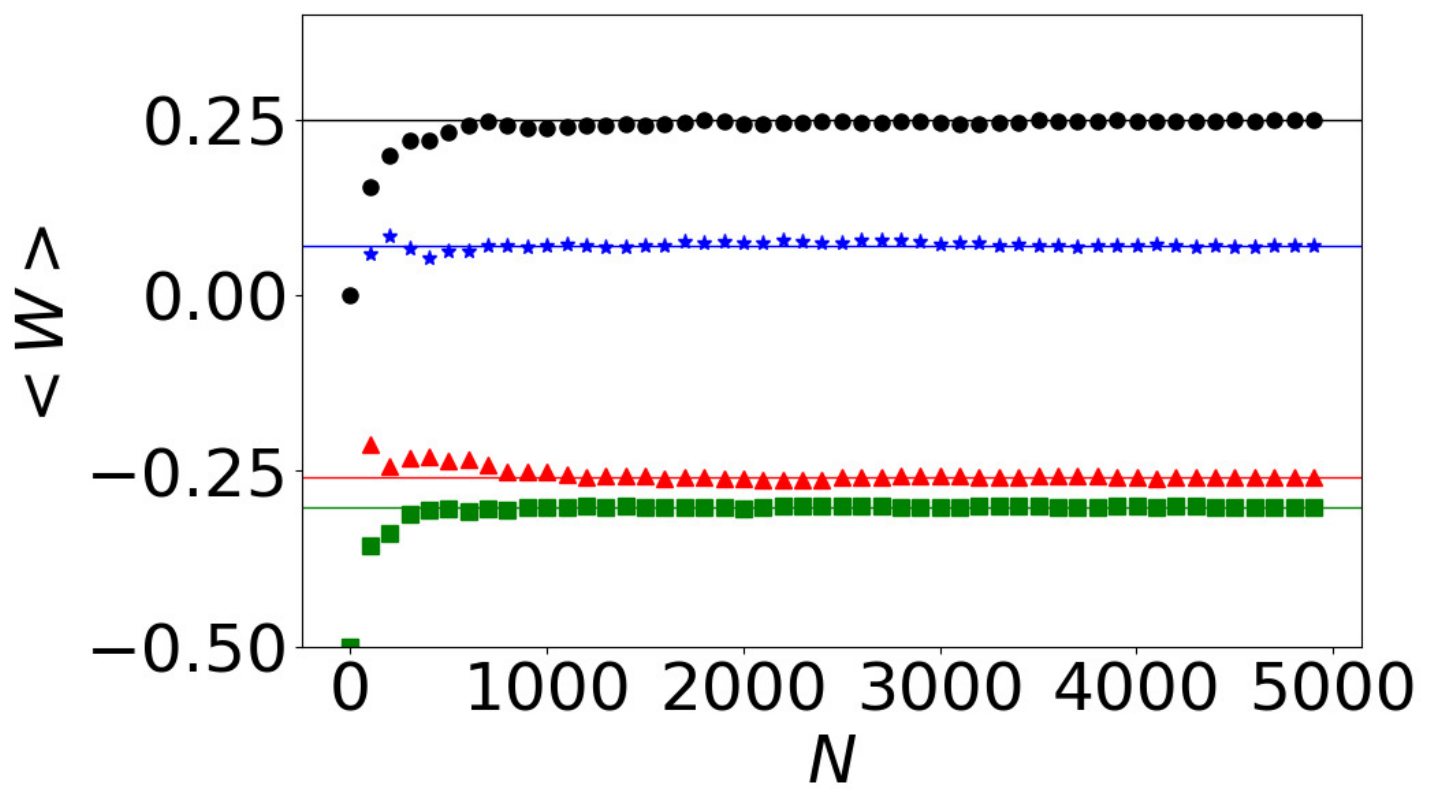}   \label{fig4:a}}
		\quad
		\subfloat[]{\includegraphics[width=0.45\linewidth]{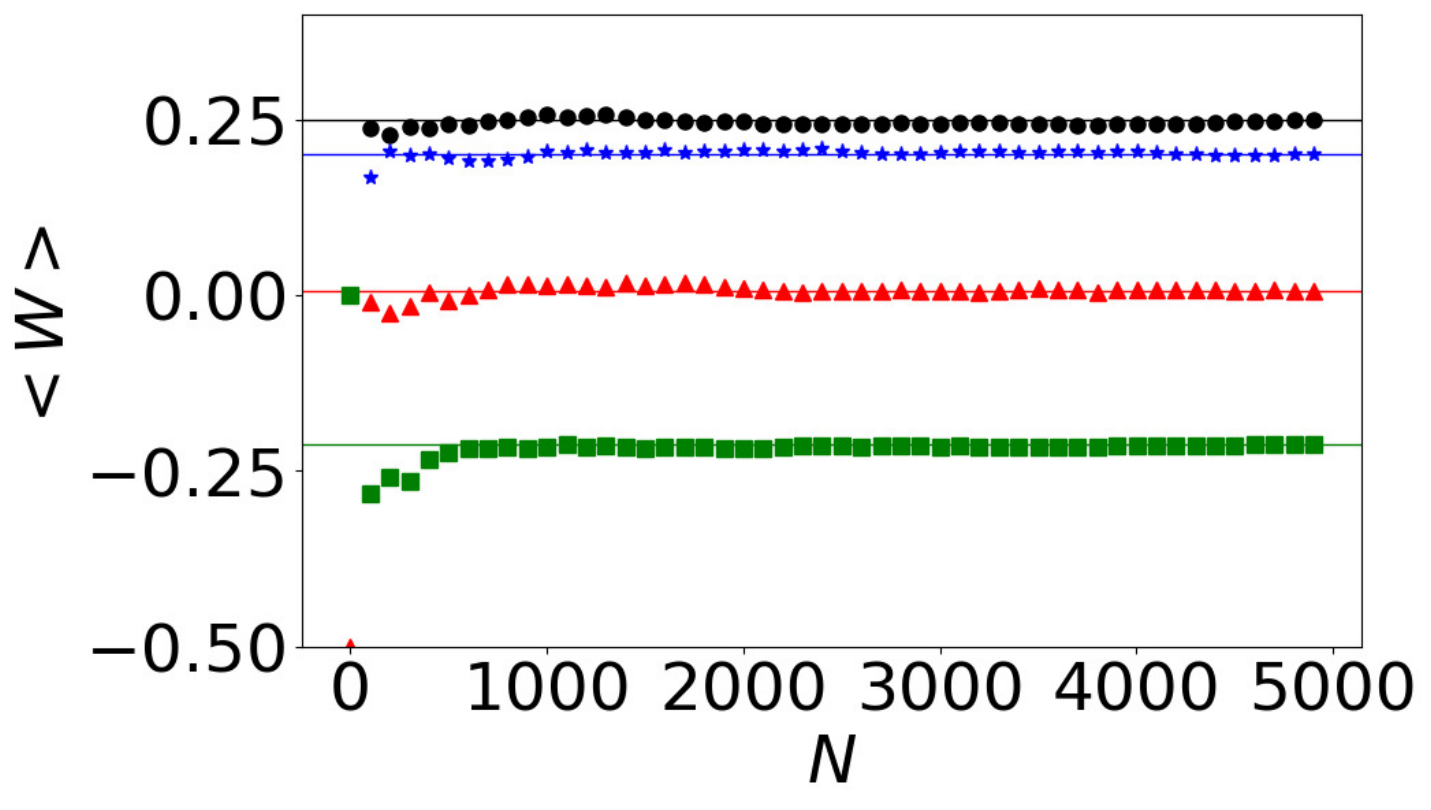}   \label{fig4:b}}
		\hfil
		\subfloat[]{\includegraphics[width=0.45\linewidth]{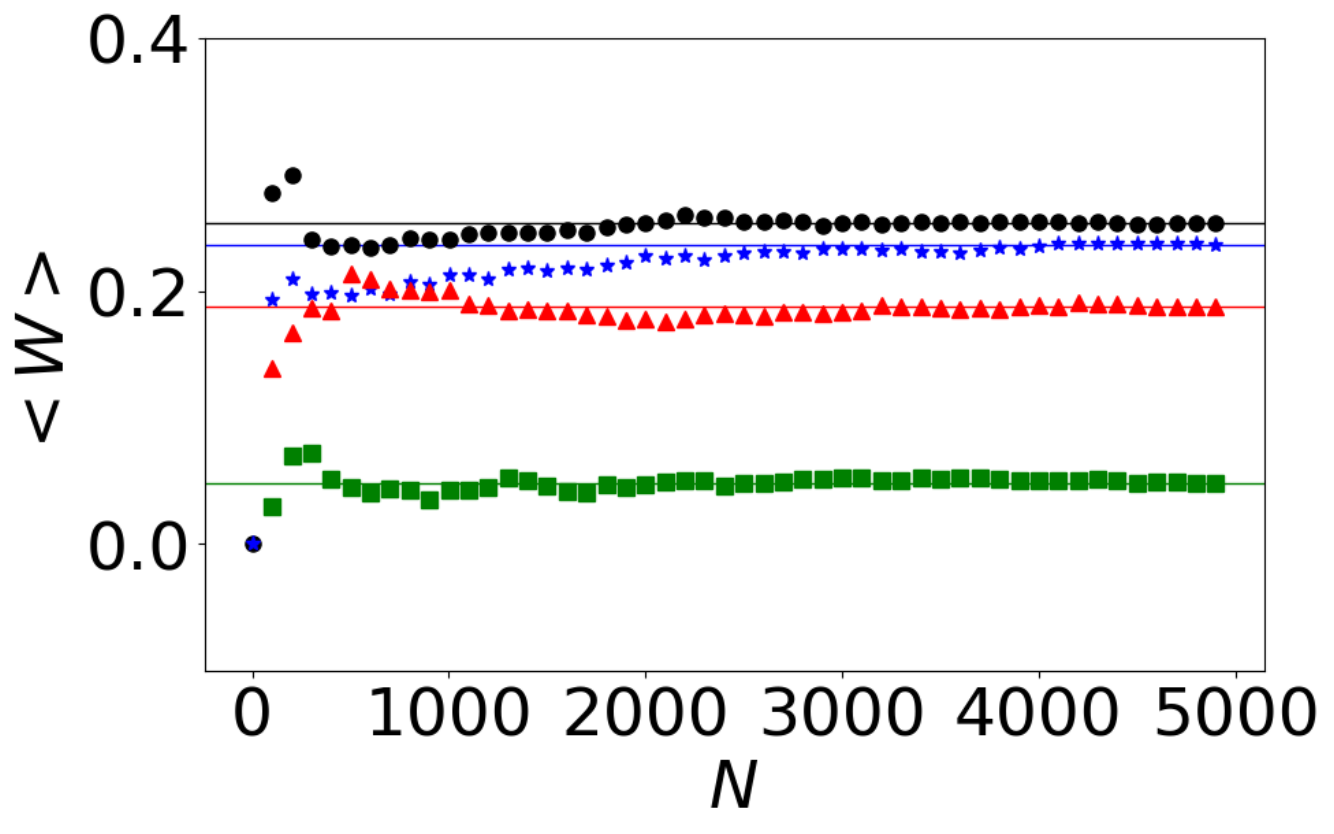}   \label{fig4:c}}
		 \caption{Results for the expectation values $\langle W \rangle$ of the Bell state witness given in Eq.~(\ref{eq10}) with respect to the number of runs $N$ and for different values of the human vision threshold $K$. The black circles, the blue star, the red triangle and the green square represent $K=1$, $K=3$, $K=5$, $K=7$ respectively. The markers on the graph are represented every hundred steps. For each threshold, the horizontal solid lines represent the last value of the witness expectation values. The probability of seeing is evaluated for $\eta=0.08$. In these graphs a human eye and an ideal photodetector have been used. $\langle W \rangle$ for the additive noise (a) $D=0$, (b) $D=2$ and (c) $D=5$.}
		 \label{Fig4}
\end{figure*}

\begin{figure*}[t!]
		\centering
  \captionsetup{justification=justified}
		\subfloat[]{\includegraphics[width=0.45\linewidth]{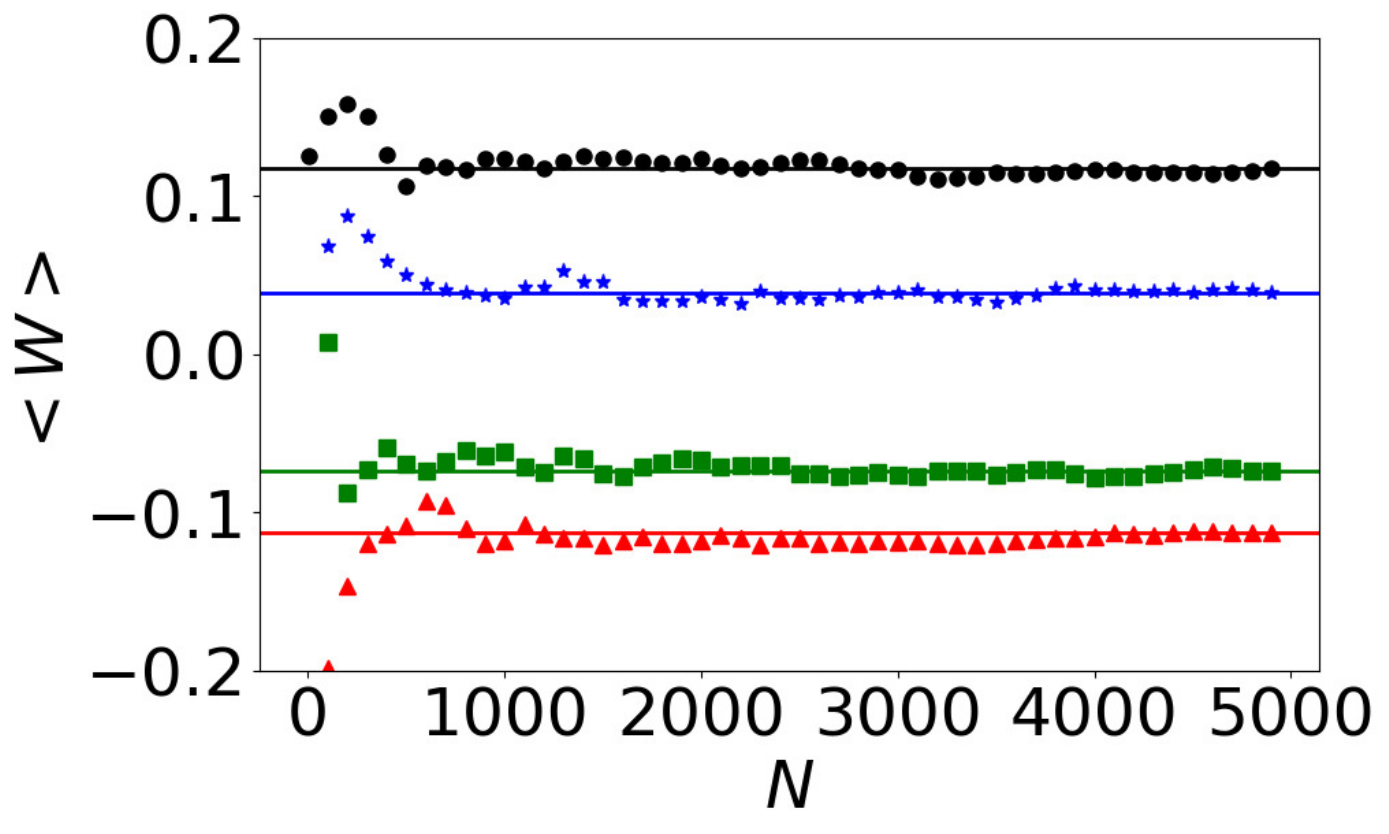}  \label{fig5:a}}
		\quad
		\subfloat[]{\includegraphics[width=0.45\linewidth]{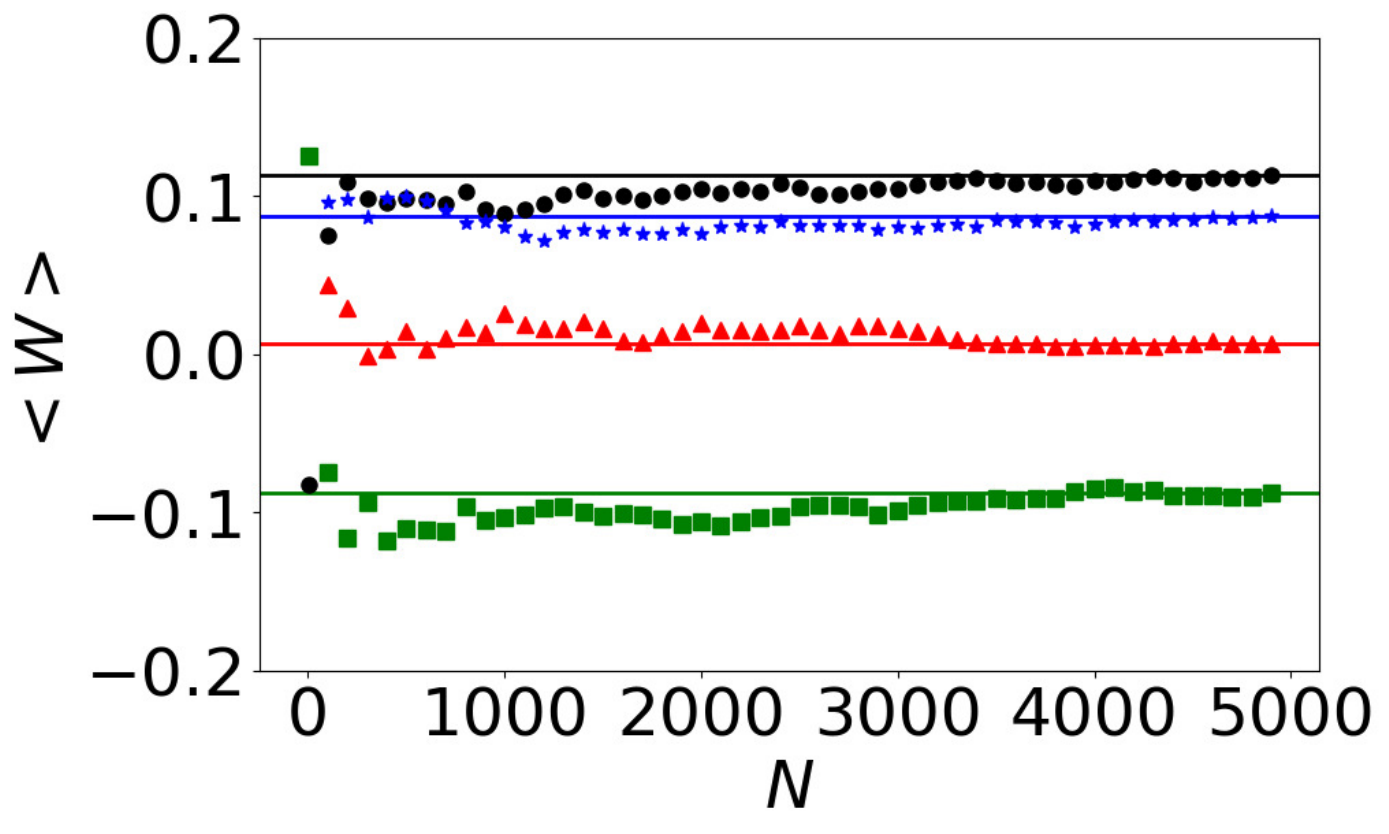}  \label{fig5:b}}
		\hfil
		\subfloat[]{\includegraphics[width=0.45\linewidth]{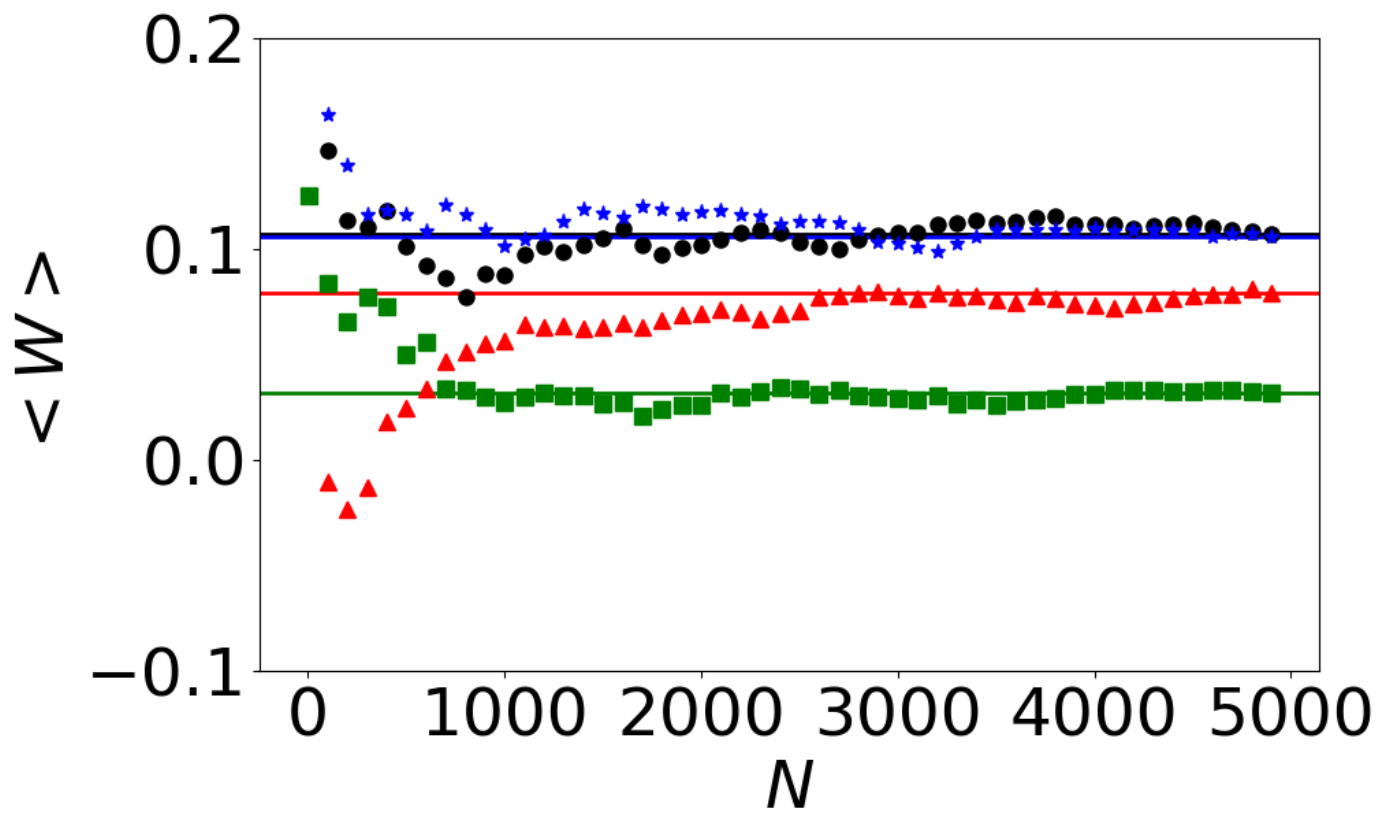}  \label{fig5:c}}
		 \caption{Results for the expectation values $\langle W \rangle$ of the W state witness given in Eq.~(\ref{eq11}) with respect to the number of runs $N$ and for different values of the human vision threshold $K$. The black circles, the blue star, the red triangle and the green square represent $K=1$, $K=3$, $K=5$, $K=7$ respectively. The markers on the graph are represented every hundred steps. For each threshold, the horizontal solid lines represent the last value of the witness expectation values. The probability of seeing is evaluated for $\eta=0.08$. In these graphs a human eye and an two ideal photodetectors have been used. $\langle W \rangle$ for the additive noise (a) $D=0$, (b) $D=2$ and (c) $D=5$.}
		 \label{Fig5}
\end{figure*}

To have a more faithful simulation, we consider an additive noise in our study.
We use a generic model, in which the parameter $D$ is added to the average photon number $|\alpha|^2$ appearing in the probability of seeing. $D$ represents the dark noise which takes into account false positive responses by the subjects. Since there is no experimental data for the parameter $D$ in such low photon number, we consider arbitrary values between 0 and 5. Introducing noise has the potential to diminish the distinction between states with and without a photon. As illustrated in~\ref{Fig3}, it becomes evident that for both larger and smaller values of $\alpha$, the probabilities associated with the presence and absence of a photon converge, diminishing their distinguishability. The simulation is run $N$ times to calculate the probabilities needed. This means that we have sent the entangled or multipartite state $N$ times and we have received the response of either the human eye or the photodetector.

While our work primarily explores the noise scenarios unique to psychophysical experimental conditions and their effects, it is worth noting that error correction techniques have been a significant area of research in quantum information science~\cite{devitt2013quantum,sivak2023real}. These techniques aim to correct measurement operator errors and enhance the reliability of quantum systems.
%%%%%%%%%%%%%%%%%%%%%%%%%%%%%%%%%%%%%%%%%%%%%%%%%%%%%%%%%%%

%%%%%%%%%%%%%%%%%%%%%%%%%%%%%%%%%%%
\section{Results}
\label{sec:results}

In the case of one human eye and an ideal photodetector, the expectation values of the entanglement witness, for the bipartite case and for different thresholds according to the number of runs of the simulation, are plotted as shown in~\ref{Fig4}. We also investigate the effect of additive noise in the probability of seeing on our results.

When looking at~\ref{fig4:a}, we see that for $K$ superior to 4 we get a negative value for the witness. So, one human eye and an ideal photodetector can be used to detect entangled state in the case of bipartite entanglement. For $K$ inferior to 4, the witness is positive. This might be due to the fact that the displacement operator has been optimized for $K=7$.
When adding noise, as shown in~\ref{fig4:b} and~\ref{fig4:c}, the entanglement detection becomes less probable according to the amount of noise. For $D=2$, the detection is possible only for $K=7$ and becomes impossible for $D \geq 5$.
When considering two human eyes for the detection of Bell state, entanglement detection is possible without noise and for additive noise corresponding to $D<3$.

In~\ref{Fig5} we consider the case of one human eye as photodetector and two ideal photodetectors. The results are very similar to the Bell state case. The multipartite detection is possible if we consider an additive noise strictly inferior to $D=5$. However, when considering two human eyes and one ideal photodetector, or three human eyes, the detection of multipartite entanglement becomes critical. For two human eyes and one ideal photodetector, the multipartite detection is possible only for limited cases, that is, for a noise strictly inferior to $D=2$, and only for a threshold $K=5$. For three human eyes, the value of the witness is positive regardless of the value of the threshold and without additive noise.

The feasibility of using humans as photodetectors to resolve single photons and entangled pairs of photons have already been shown in experiments in the literature. Our scheme do not improve the detection efficiency of humans in this regard, but contributes partly to answer the critical question of  what are the limits of human photodetectors to detect higher dimensional entanglement in realistic psychophysical experimental conditions. We find that the additional noise in such experiments seriously hinder the resolution of even a tripartite entangled state and we determine how the experiments should be designed with combination of coherent amplifiers, artificial detectors and humans to determine the tripartite entanglement with acceptable success rates.

\section{Amplification methods}
\label{sec:amp}

%==========================================================
%\begin{figure*}[t!]
%        \captionsetup{justification=justified}
%	\centering
%		\subfloat[]{\includegraphics[width=0.45\linewidth]{img6.1.eps}  \label{fig6:a}}
%		\quad
%		\subfloat[]{\includegraphics[width=0.45\linewidth]{img6.2.eps}  \label{fig6:b}}
%  		\hfil
%		\subfloat[]{\includegraphics[width=0.45\linewidth]{img6.5.eps}  \label{fig6:c}}
%		\quad
%		\subfloat[]{\includegraphics[width=0.45\linewidth]{img6.6.eps}  \label{fig6:d}}
%	 \caption{\textcolor{blue}{The mean photon number is shown by the solid green curve, while the variance and the standard deviation of the photon number are respectively shown by the dashed red curve and the dashed-dotted blue curve. (a) The displacement operator is applied to $\ket{0}$ and the graphs are plotted according to the modulus of the displacement (coherence) parameter $r$. (b) The displacement operator is applied to $\ket{1}$ and the graphs are plotted according to the displacement parameter $r$. (c) Amplitude squeezing is applied to $\ket{0}$, according to $r$. (d) Amplitude squeezing is applied to $\ket{0}$, according to $r$. For (c) and (d), the modulus and the phase of the squeezing parameter are $s=0.5$ and $\theta=0$ respectively and the phase of the displacement parameter is taken as $\phi = 0$.}}
%	 \label{Fig6}
%\end{figure*}
%%%%%%%%%%%%%%%%%%%%%%%%%%%%%%%%%%%%%%%%%%%

We have seen that an amplification scheme is necessary to increase the average photon number pre-measurement above the detection threshold. Previous studies have achieved this through coherent amplification techniques, such as the displacement operator~\cite{Vivoli:16}, stimulated emission~\cite{PhysRevLett.103.113601}, and optical parametric amplification~\cite{spagnolo2011hybrid,pakarzadeh2021thermally}. However, these methods also result in an increase in photon number fluctuations. It actually also increases the dark count making the distinguishability between photon and no photon more difficult. To minimize this noise, a natural direction would be the use of amplitude squeezing. The application of amplitude squeezing~\cite{PhysRevA.53.1096} utilizes both the displacement and squeezing operators such that the amplitude squeezing operator acting on the input state can be written as
\begin{align}
\ket{n,\alpha,z}&=\hat{D}(\alpha)\hat{S}(z)\ket{n},\\
\hat{D}(\alpha) &= \mathrm{e}^{\alpha \hat{a}^{\dag} - \alpha^{*} \hat{a}},\\
\hat{S}(z) &= \mathrm{e}^{\frac{1}{2}(z^{*} \hat{a}^{2} - z \hat{a}^{\dag 2})}.
\end{align}
Here the displacement parameter $\alpha =re^{i\phi}$ and the squeezing parameter $z=se^{i\theta}$ are complex numbers. $\hat{a}^{\dag}$ and $\hat{a}$ are respectively the creation and annihilation operators. Amplitude squeezing appears for specific values of the phases $\phi$ and $\theta$~\cite{PhysRevA.53.1096}. The photon number fluctuations reduce at the cost of increasing phase noise, consistent with the phase-number Heisenberg uncertainty relation. The general expression for mean photon number $\bar{N}$ and the photon number variance $V(\hat{N})$ are given by
\begin{align}
\langle \hat{N} \rangle&  = \langle \hat{a}^{\dag} \hat{a} \rangle,\\
V(\hat{N})&= \langle \hat{N}^{2} \rangle - \langle \hat{N} \rangle ^{2},
\end{align}
with $\hat{N}=\hat{a}^{\dag} \hat{a}$.
The expression of the mean photon number and the photon number variance can be derived for amplitude squeezed state as follows~\cite{dantas1998statistical},
\begin{align}
    \langle \hat{N} \rangle & = |\alpha|^2 + (2n +1)\sinh^2(s) + n,\label{eq30}\\
    \begin{split}
        V(\hat{N})&= |\alpha|^2 (2n+1) [ \mathrm{e}^{-2s} \cos^2 (\theta-\frac{\phi}{2})\\
        &+ \mathrm{e}^{2s}  \sin^2 (\theta-\frac{\phi}{2}) ]\\
        &+ (n^2 + n + 1)2\sinh^2(s)\cosh^2(s).\label{eq31}
    \end{split}
\end{align}

\begin{figure}[t!]
        \centering
        \captionsetup{justification=justified}
        \includegraphics[width=\linewidth]{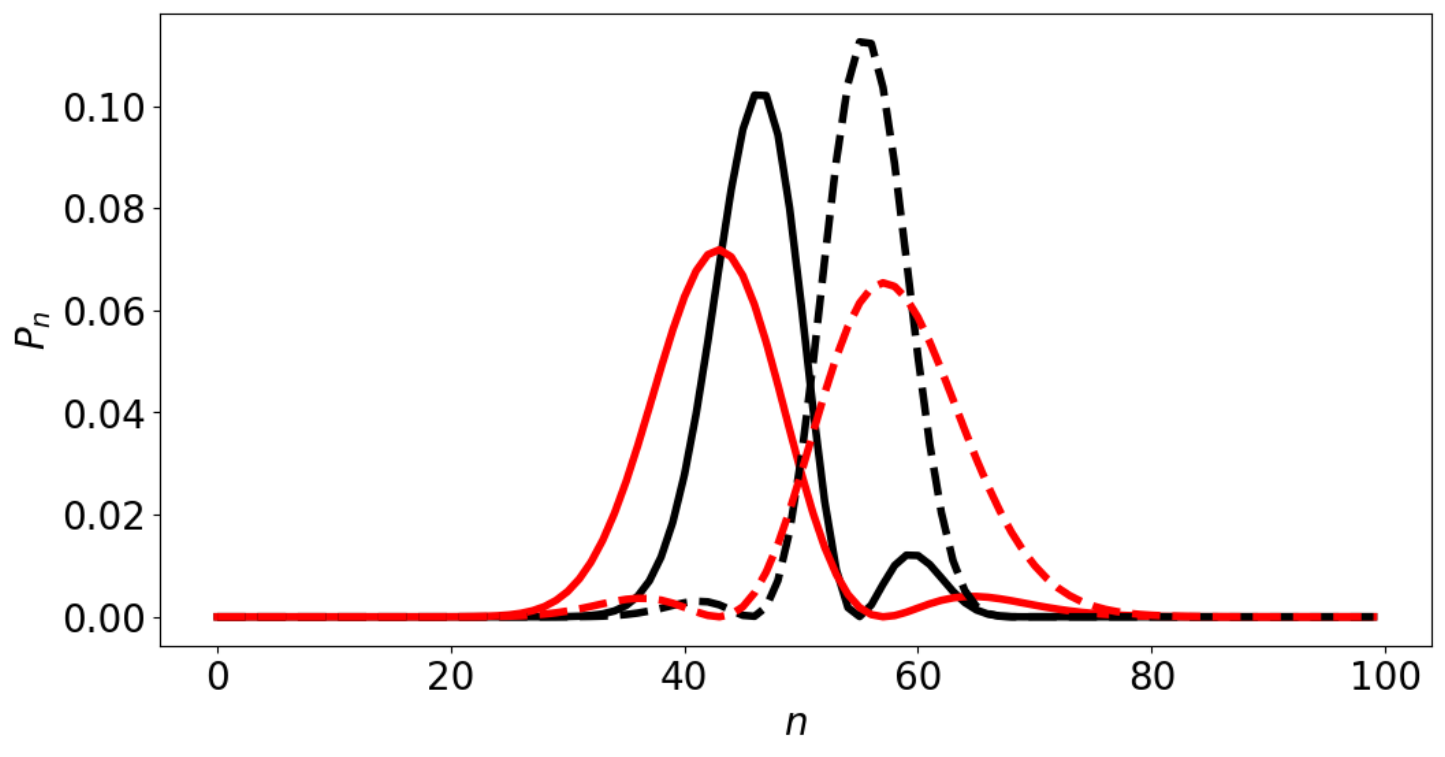}
        \caption{Probability distribution $P_n$ according to the photon number $n$. The red dotted curve corresponds to the case of applying a displacement operator to $\ket{0'}$. The solid red curve corresponds to the case of applying a displacement operator to $\ket{1'}$ . The black dotted curve corresponds to the case of applying amplitude squeezing to $\ket{0'}$ and the solid red curve corresponds to the case of applying amplitude squeezing to $\ket{1'}$. The modulus of the displacement (coherence) parameter is taken as $r=\sqrt{50}$. The modulus and the phase of the squeezing parameter are $s=0.5$,  and $\theta=0$, respectively. The phase of the displacement parameter is taken as $\phi = 0 $.}
       \label{Fig7}
\end{figure}

Opting for amplitude squeezing over displacement does not enhance the distinguishability between with and without photons (see~\ref{Fig7}). Although it reduces variance, it concurrently amplifies dark counts, making it an unreliable choice for amplification and complicating the process of entanglement witnessing.

In the realm of quantum optics, researchers have explored the concept of an ideal photon number amplifier, aiming to convert an input state with `$n$' photons to an output state with `$G$' photons (where `$G$' $\textgreater$ 1). Both linear and non-linear amplification methods have been explored \cite{ho1994scheme,d1992hamiltonians,caves2012quantum,propp2019nonlinear}, with non-linear amplification showing promise for achieving an ideal photon number amplifier.

The standard approach to describing linear phase-preserving quantum amplification of a bosonic mode amplitude `$a$' is through Caves' relation~\cite{propp2019nonlinear},
\begin{equation}
\hat{a}_{\mathrm{out}} = \sqrt{G} \hat{a}_{\mathrm{in}} + \sqrt{G - 1} \hat{b}^+_{\mathrm{in}},
\end{equation}
where $\hat{a}^{\dag}$ and $\hat{a}$ are respectively the creation and annihilation operators of mode `$a$', and $\hat{b}^{\dag}$ and $\hat{b}$ are respectively the creation and annihilation operators of mode `$b$'.
This relation introduces an output field, a combination of the input field amplitude multiplied by $\sqrt{G}$ and an additional term related to an auxiliary mode `$b$'. The presence of mode `$b$' introduces extra noise, even if it is in a vacuum state, affecting the output field's characteristics. The resulting variance expression reflecting the impact of this extra noise on the amplified field is
\begin{align}
\sigma^2_{(\hat{a}^{\dag}\hat{a})_\mathrm{out}} =& G\Delta n_a^2 +(G-1)^2\Delta n_b^2 \\&+ G(G-1)(2\bar{n}_a\bar{n}_b +\bar{n}_a+\bar{n}_b+1),
\end{align}
where $\bar{n}_a=\langle (\hat{a}^{\dag}\hat{a})_\mathrm{in}\rangle$ and
$\bar{n}_a^2 +\Delta \bar{n}_a^2=\langle(\hat{a}^{\dag}\hat{a})^2_\mathrm{in}\rangle$. We obtain similar expressions for mode `$b$'.
In the Heisenberg picture, non-linear transformations of the bosonic annihilation operator can be achieved while maintaining bosonic commutation relations. In the ideal case, coupling to a reservoir mode `$b$' resulting in an output mode with no additional amplification noise is given by
\begin{equation}
\hat{b}_{\mathrm{out}}^{\dag}\hat{b}_{\mathrm{out}}=\hat{b}_{\mathrm{in}}^{\dag}\hat{b}_{\mathrm{in}}+G \hat{a}_{\mathrm{in}}^{\dag}\hat{b}_{\mathrm{in}},
\end{equation}
which gives as variance
\begin{equation}
\sigma^2_{(\hat{b}^{\dag}\hat{b})\mathrm{out}} = \Delta n_b^2 + G^2\Delta n_a^2.
\end{equation}
The number fluctuations in the auxiliary mode are not amplified, particularly if the input field is in a number state with zero fluctuation. In a more realistic scenario where reservoirs consist of many modes (for example `$G$' modes), the number fluctuations in the output mode are a combination of the amplified fluctuations from both the input and reservoir modes
\begin{equation}
\sigma^2_{(\hat{b}^{\dag}\hat{b})\mathrm{out}} = G\Delta n_b^2 + G^2\Delta n_a^2.
\end{equation}
The amplification of noise is a common feature in both linear and non-linear amplification methods. However, the non-linear approach has the advantage of conditional coupling, which can reduce noise under certain conditions. Envisaging a superior amplification method compared to displacing the state or using amplitude squeezing is crucial for achieving better distinguishability between the one-photon and zero-photon states. The key requirement for entanglement witnessing is the ability to clearly distinguish the absence of photons when they are not present and their presence when they are. Ideal photon amplifiers offer the potential for maintaining a zero-photon state with minimal noise and amplifying the one-photon state beyond a specified threshold. While the experimental implementation may pose challenges, striving for the realization of ideal photon amplifiers represents a promising direction for future advancements.

%%%%%%%%%%%%%%%%%%%%%%%%%%%%%%%%%%%%%%
\section{Discussion and Conclusion}
\label{sec:conclusion}

In our study, we investigated the conditions for detecting multipartite quantum entanglement using the human eye in psychophysical experiments. The coherent amplification technique was generalized for detecting entanglement in a bipartite entangled Bell state to a tripartite entangled W state. We utilized entanglement witnesses that could be measured using polarization analyzers in the interferometric setup. To evaluate the entanglement witnesses, we simulated human responses by modifying the probability of seeing~\cite{Vivoli:16} to include the additive noise in the retina to simulate false positives in psychophysical experiments.

When the additive noise was neglected, our results showed that the detection of the entangled photons in the Bell state was possible using one human eye and one ideal photodetector, or with two human eyes as photodetectors, for threshold values $K > 4$.  However, when the intrinsic retinal noise ($D$) was added to the simulation, the detection became more challenging but was still achievable depending on the noise level. In the case of the W state scenario, the detection was not possible with three human eyes even without noise addition, but by replacing one or two human eyes with perfect photodetectors, multipartite detection became possible. As with the Bell state, adding noise to the simulation resulted in a deterioration of the detection, but it was still feasible if the noise remained below a certain threshold.

In conclusion, the possibility of using the human eye as a photodetector to detect entanglement appears to be dependent on three critical factors, the order (dimensionality) of entanglement, $D$, and $K$. Detection of higher dimensional entanglement is more difficult by only human eyes, but it can be possible (i) if high-efficiency photodetectors are used together with human eyes; (ii) if $D$ is smaller than a few photons ($D \sim 5$); and (iii) signal is coherently and optimally amplified for a given visual threshold.  While the efficiency of the photodetector model of the human eye is another biometric factor that severely limits entanglement detection, it is kept as a constant in the present work, and its small variations were assumed to have no significant impact on the conclusion of the possibility of detection of entanglement. Guided by the ranges of influence of significant biometric factors discussed here, interrogation strategies and the design of psychophysical experiments can be optimized to use higher dimensional entangled states to probe the human visual system with the potential benefits of quantum metrology. We have been exploring ways to improve the amplification process, beginning with amplitude squeezing. However, our observations indicate that this method worsens the distinguishability between photon and non-photon states by elevating dark counts.
Given this context, we have explored various alternative amplification approaches documented in the literature. It appears that the application of non-linear amplification aligns more closely with the ideal photon amplifier scenario, presenting the potential to effectively alleviate noise-related challenges.

This study primarily aimed to explore the fascinating and fundamental aspects of utilizing the human eye for detecting quantum entanglement, similar to the pioneering work on Bell states. Beyond its scientific curiosity, this research could yield practical benefits.
Indeed, our findings on applying scaled multipartite quantum entanglement in psychophysical experiments might open doors for quantum-based approaches to investigating aspects of human psychology and vision.

Our study focused on utilizing entangled states in the horizontal (H) and vertical (V) polarization basis for photodetection.
In recent years a novel research direction has emerged in which structured light with left (L) and right (R) circularly polarized states were used for human vision experiments. In this line of work, the researchers successfully utilized R and L polarization coupled orbital angular momentum (OAM) states with an azimuthal phase dependence to trigger the entoptic vision of the human subject in which the number of visible azimuthal fringes become proportional to the difference in the OAM values~\cite{sarenac2020direct,sarenac2022human}. Utilizing OAM structured light has shown promise as humans were able to distinguish OAM states with high accuracy and it was argued that one can use such states for diagnosis of macular degeneration, as the number of observed fringes would be different between healthy and unhealthy subjects~\cite{pushin2023structured,pushin2024psychophysical}. Although using such states is an interesting research direction to investigate entanglement sensitivity of the human visual system, challenges exist in adapting such techniques to our proposed experimental scheme. To see entoptic phenomena, peripheral vision needs to be activated which usually requires a much brighter light source than what scotopic vision requires for activation. Therefore, for our scotopic vision based proposal, we choose the simpler case of H and V entangled light sources with wavelengths around the peak wavelength of human rod cell's absorption curves between 495 and 505 nm~\cite{Tinsley2016}.

Finally, the intersection of quantum mechanics and biology, particularly in the realm of vision, presents an interesting direction. Quantum biology's role in vision suggests a subtle, yet profound, connection where quantum mechanics may influence how we perceive light. This means we might view both light and the eye itself through a quantum perspective~\cite{lambert2013quantum,fleming2011quantum}. Such an understanding can contribute to technical advancements from neuroscience to medicine, in which the visual system is the subject of investigation. Quantum tomography is an indispensable tool of quantum sciences and technologies~\cite{MAURODARIANO2003205,Cramer2010,Heinosaari2013,PhysRevA.77.032322,PhysRevLett.90.193601,Lundeen2009,pedram2024nonlocality} which can bridge this gap. A recent study showed that quantum process tomography methods can be applied to the human visual system to reconstruct the visual perception model using Bayesian statistics~\cite{vanderReep:23}. In this study, light pulses with Poissonian photon statistics were used. Using quantum entangled sources of light it might be possible to come up with, and validate, a finer perception model using fewer number of trials.

\section*{Acknowledgements}

This work is based upon research supported by the Scientific and Technological Research Council of Turkey (T\"UBITAK), grant No. 120F200. We thank Professor Igor Meglinski from Aston University for insightful discussions.

\bibliography{references}

\end{document}